\documentclass[conference]{IEEEtran}
\IEEEoverridecommandlockouts
\usepackage{cite}
\usepackage{balance}
\usepackage{amsmath,amssymb,amsfonts}
\usepackage{algorithmic}
\usepackage{graphicx}
\usepackage{textcomp}
\usepackage{xcolor}
\def\BibTeX{{\rm B\kern-.05em{\sc i\kern-.025em b}\kern-.08em
    T\kern-.1667em\lower.7ex\hbox{E}\kern-.125emX}}

\usepackage{color}
\usepackage{comment}
\usepackage{caption}
\usepackage{subcaption}

\usepackage{amsmath}
\usepackage[export]{adjustbox}

\usepackage{ifpdf}
\ifpdf
    \graphicspath{{./figs/}}
\else
    \graphicspath{{./figs/ps/}}
\fi

\ifCLASSOPTIONcompsoc
  \usepackage[nocompress]{cite}
\else
  \usepackage{cite}
\fi
\ifCLASSINFOpdf
\else
\fi
\begin{document}



\title{AI Aided Noise Processing of Spintronic Based IoT Sensor for Magnetocardiography Application}


%
%
%
%
\author{
	\IEEEauthorblockN{
	Attayeb Mohsen\IEEEauthorrefmark{1}$^1$,
	Muftah Al-Mahdawi\IEEEauthorrefmark{2}$^2$,  
	Mostafa M. Fouda\IEEEauthorrefmark{3}$^3$,\\
	Mikihiko Oogane\IEEEauthorrefmark{2}\IEEEauthorrefmark{4}$^4$,
	Yasuo Ando\IEEEauthorrefmark{2}\IEEEauthorrefmark{4}$^5$,
	and Zubair Md Fadlullah\IEEEauthorrefmark{5}$^6$.
	}
	
	    \IEEEauthorblockA{\IEEEauthorrefmark{1}Artificial Intelligence Center for Health and Biomedical Research (ArCHER),\\ National Institutes of Biomedical Innovation, Health and Nutrition (NIBIOHN), Osaka, Japan.
		}
	    \IEEEauthorblockA{\IEEEauthorrefmark{2}Center for Science and Innovation in Spintronics (Core Research Cluster), Tohoku University, Sendai, Japan. 
	    }
    	\IEEEauthorblockA{\IEEEauthorrefmark{3}Department of Electrical and Computer Engineering, Tennessee Tech University, Cookeville, TN, USA, and \\ Faculty of Engineering at Shoubra, Benha University, Egypt. 
		}
		\IEEEauthorblockA{\IEEEauthorrefmark{4}Department of Applied Physics, Graduate School of Engineering, Tohoku University, Sendai, Japan 
	    }
		\IEEEauthorblockA{\IEEEauthorrefmark{5}Computer Science Department, Lakehead University, and \\
		Thunder Bay Regional Health Research Institute (TBRHRI), Thunder Bay, Ontario, Canada. \\
		Emails: $^1$attayeb@nibiohn.go.jp, $^2$mahdawi@mlab.apph.tohoku.ac.jp,\\
		$^3$mfouda@ieee.org, 
		$^4$oogane@mlab.apph.tohoku.ac.jp, 
		$^5$ando@mlab.apph.tohoku.ac.jp,\\
		$^6$\{zubair.fadlullah@lakeheadu.ca, fadlullz@tbh.net\}.
		}
}

\maketitle
\begin{abstract}
As we are about to embark upon the highly hyped ``Society 5.0'', powered by the Internet of Things (IoT), traditional ways to monitor human heart signals for tracking cardio-vascular conditions are challenging, particularly in remote healthcare settings. 
On the merits of low power consumption, portability, and non-intrusiveness, there are no suitable IoT solutions that can provide information comparable to the conventional Electrocardiography (ECG). In this paper, we propose an IoT device utilizing a spintronic-technology-based ultra-sensitive Magnetic Tunnel Junction (MTJ) sensor that measures the magnetic fields produced by cardio-vascular electromagnetic activity, \emph{i.e.}~Magentocardiography (MCG). We treat the low-frequency noise generated by the sensor, which is also a challenge for most other sensors dealing with low-frequency bio-magnetic signals. Instead of relying on generic signal processing techniques such as moving average, we employ deep-learning training on bio-magnetic signals. Using an existing dataset of ECG records, MCG signals are synthesized. A unique deep learning model, composed of a one-dimensional convolution layer, Gated Recurrent Unit (GRU) layer, and a fully-connected neural layer, is trained using the labeled data moving through a striding window, which is able to smartly capture and eliminate the noise features. Simulation results are reported to evaluate the effectiveness of the proposed method that demonstrates encouraging performance. 

\end{abstract}

\section{Introduction}

Despite the recent proliferation of Internet of Things (IoT) sensors and wearable technologies, heart monitoring at non-clinical, remote settings (e.g., at home) over prolonged period is challenging and not accurate compared to the baseline ElectroCardioGraphy (ECG) method. As a non-intrusive alternative to the widely available ECG, MagentoCardioGraphy (MCG)~\cite{tavarozzi_2002} appeared as a promising technique, which measures the magnetic field produced by electrical activity in the human heart. 
Although MCG measurement possesses clinically useful information for a more accurate and advanced cardiac diagnosis~\cite{Haberkorn2006PseudoCD},  
it is not without shortcomings. The multi-channel MCG equipment usually comprise more than 50 detectors called Superconducting Quantum Interference Devices (SQUIDs), which requires expensive liquid helium cooling~\cite{smith_2006}. 
Furthermore, due to the expense and need for magnetic shielding for SQUID based MCG~\cite{MASLENNIKOV201288}, its diagnostic usefulness compared to ECG devices is questioned in~\cite{KONG2012286}.

\begin{figure}[t]
	\includegraphics[width=\columnwidth]{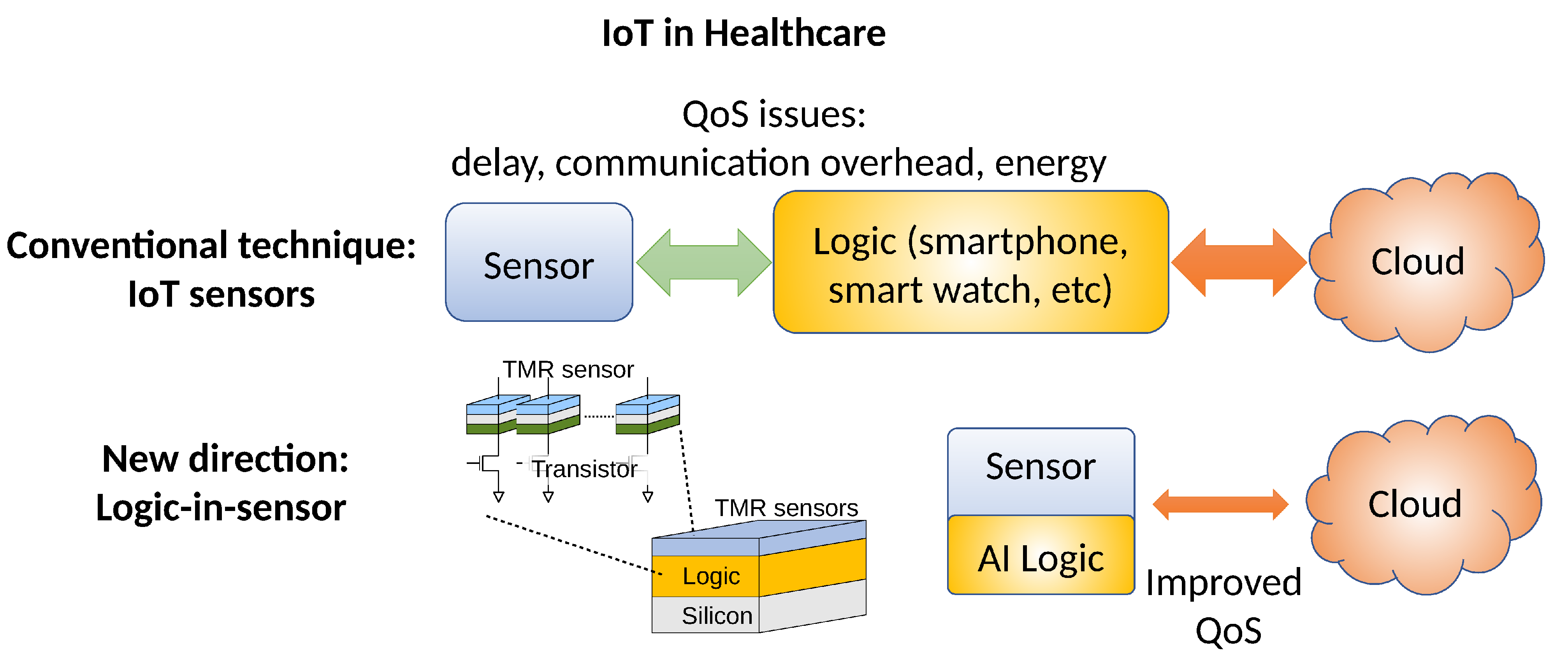}
	\caption{Envisioned logic-in-sensor architecture where spintronic/other sensors integrated with Artificial Intelligence (AI) logic can revolutionize IoT-based smart health applications.
	}
	\vspace{-0.4cm}
	\label{fig-archi}
\end{figure}

In this paper, we present a novel spintronic sensor based on the Tunneling MagnetoResistance (TMR) effect in Magnetic Tunnel Junction (MTJ) devices developed by the physicists of our research group that has great potential for sensing human heart and brain signals \cite{Fujiwara_2018}. The sensor is small and it can be fused on-chip with silicon logic circuits to construct portable lab/logic-in-sensor diagnostics devices as depicted in~Fig.~\ref{fig-archi}. Due to this logic-in-sensor architecture, it improves the Quality of Service (QoS) issues such as communication delay, overhead, and energy expenditure, and therefore, can be used for long-term, accurate monitoring of cardiac cycle of the patient without interfering with the subject's day-to-day activities. 

One of the major challenges for biomagnetic detection is the sensor noise at low frequencies, usually represented as $1/f$ noise, where $f$ is the spectral frequency \cite{lei_2011}. This creates a unique problem for measuring MCG signals using the spintronic sensor simply because the heart also oscillates at low frequency, producing signals in the same frequency band as the noise.
To tackle this problem, in this paper, we  propose a deep learning-based approach (an Artificial Intelligence (AI) methodology) to perform noise processing on MCG signal, which is synthesized from a public ECG dataset to replicate the properties of our spintronic MTJ sensors. The proposed deep learning-based method takes advantage of a uniquely constructed structure combining a one-dimensional (1-D) convolution layer, a Gated Recurrent Unit (GRU) layer, and a fully-connected (feed forward) neural network layer. The 1-D convolution layer performs the automated feature extraction from the MCG signal, and passes the feature sets to a GRU layer. The GRU layer performs a non-linear mapping on the resulting feature map based on past features. 
The output of the GRU layer is formatted (i.e., flattened) to serve as the input to a fully-connected layer, which generates the ECG signal corresponding to the MCG input signal. Using the MCG-ECG labels, the fully-connected network is trained to improve the prediction accuracy. The performance of our proposed deep learning method is validated by conducting computer-based simulations. The trained model can be easily incorporated into the sensor node incorporating our conceptualized spintronic IoT sensor for MCG signal monitoring.


The remainder of this paper is organized as follows. Section~\ref{related-works} presents the relevant research work. Section~\ref{system-model} describes the considered model of our proposed spintronic sensor. It also delineates how it is able to measure MCG signal. Section~\ref{problem-formulation} presents a formal problem formulation. discusses why standard filtering methods, such as the moving-average technique, are not adequate to process the MCG signal sensed by our spintronic sensor. In Section~\ref{proposal}, we present a novel deep learning model to distinguish the noise from the actual MCG signal. The performance of our proposal is validated in Section~\ref{performance}. Finally, the paper is concluded in Section~\ref{conclusion}. 



\section{Related Work}
\label{related-works}

Liu~\textit{et al.}~\cite{liu2016} surveyed extensive sensing capabilities of various spintronic sensor types for various applications ranging from the smart grid to smart healthcare. Spintronic sensors were identified to be a potential enabler to build effective Point-of-Care (POC) platform, combining smart and portable bio-sensors, computing, networking, and ICT technologies, for pervasive healthcare devices to reduce healthcare costs and improve diagnostic and monitoring efficiency, particularly in countries with large populations or rural areas. Handheld, battery-powered POC devices comprising spintronic bio-detection sensors were developed that can be exploited for detecting bio-molecules. 
On the other hand, in the research work conducted by Fujiwara \textit{et al.}~\cite{Fujiwara_2018}, an ultra-high sensitive TMR sensor was developed to measure both heart's MagnetoCardioGraphy and brain's MagnetoEncephaloGraphy (MEG) signals at the room temperature. The MCG R wave was detected without averaging, and the QRS complex was observed with a good signal to noise ratio (SNR) by averaging signals for several tens of seconds. The MCG mapping was demonstrated with high spatial resolution. Real-time MCG measurements with high spatial resolution at the room temperature may lead to significant improvements in the diagnosis of heart disease. In addition, the world-first detection by room-temperature sensors of the human brain MEG was demonstrated. The MEG alpha wave at approximately 10 Hz was observed. This demonstration of MEG measurements at room temperature is a very significant advance in biomagnetometry. The present findings are expected to lead to various room temperature biomedical applications using TMR sensors.

\section{Overview of Proposed Spintronic IoT Sensor}
\label{system-model}

Among the various types of sensors, the magnetic field sensors contribute only to 9\% of sensor shipments by Japanese companies~\cite{jeita_report}.
Among those magnetic sensors, the sensors based on Hall and anisotropic mangetoresistance effects are predominant, and TMR-based ones are the least available~\cite{fmi_report}.
 However, a shifting landscape of application demands in automotive, smart healthcare, infrastructure, and power-grid sectors will require specifications found only in TMR technology. Furthermore, magnetic sensing can be expanded to other sensing domains, such as temperature, position, pressure, acceleration, and so forth. Therefore, there is a strong potential for TMR sensors to take from the market share of other sensors, and expand into applications not possible before.
 
 For example, by developing ultra-sensitive TMR sensors, non-invasive measurements of human body biomagnetic signals were demonstrated. 
The biomagnetic fields are on an order of 1 femtotesla to 10 picotesla. Therefore, while keeping very high sensitivity of our developed IoT sensor, the noise at low frequency is crucial. Multiple approaches at the physical layer were used: integrating large arrays of MTJs ($100 \times 100$ elements) \cite{fujiwara_2013}, optimization of sensing layer materials and fabrication process \cite{fujiwara_2012,kato_2013}, decrease of MTJs resistance-area product, and signal filtering and conditioning. Recently, a significantly small detectivity of $10 pT/\sqrt{\text{Hz}}$ at 3 Hz was achieved by an industrial collaborator, which was demonstrated for MCG and MEG acquisition \cite{kumagai_2019}. 

At low frequency, the noise in our MTJ based sensor is dominated by $1/f$ character, which is a general phenomenon in various types of systems and sensors \cite{hooge_1981}. The problem is exacerbated at the high sensitivity region~\cite{wisniowski_2008}. The power spectral density (PSD) of low-frequency noise can be represented as \cite{lei_2011}:
\begin{equation}
    S_v \propto \frac{\chi}{M_s V} \frac{1}{f^\beta},
\end{equation}
where $\chi$, $M_s$, $V$, $f$, and $\beta$ denote the sensor susceptibility which is proportional to sensitivity, the sensor saturation magnetization, sensor volume, spectral frequency, and the exponent of noise spectrum, respectively.
Mitigating the $1/f$ noise in the MTJ sensor will open the way to exploit its high sensitivity and low total ownership cost compared to other magnetic/non-magnetic sensors.

\section{Problem Formulation}
\label{problem-formulation}

Remote/home monitoring of cardiac conditions using IoT sensors and wearable devices for prolonged periods has recently emerged as a hot research topic. This is because of its ability to capture the variability of the cardiac signals for a long duration to indicate irregularities of heart conditions and other related  disorders for medical analytics. Conventional ECG at the care-giving facility cannot be used for home monitoring due to technical challenges such as placing the electrodes on the patient body and managing the ECG equipment at home. On the other hand, Holter monitors for collecting ECG data~\cite{holter} can be used for home monitoring. However, Holter monitors are expensive, have to be worn (i.e., interferes with the daily activity of the user), and are usually rented to the patient for limited time of use (typically for a day). Also, the care-givers need to download and analyze the collected ECG data from the Holter monitor which involves typically a few days. The high cost of these devices prevents long term monitoring use like athletic performance of sportsmen, elderly patients aging at home, patients taking medications causing side-effects, and so forth~\cite{holter}. Therefore, we explore the potential of our affordable spintronic sensor to capture MCG information in a non-intrusive manner and aim to solve the cardiac source imaging problem efficiently by mapping the obtained MCG trace to find the corresponding ECG signal. 

While ECG provides adequate information regarding normally-oriented and posterior sources, MCG is able to provide a higher resolution of information by characterizing tangential anterior sources of the human heart. However, employing spintronic sensors to detect the magnetic field of the heart with high sensitivity requires noise filtering. In order to filter the intrinsic noise of the considered spintronic IoT sensor, using the moving average technique (the most widely adopted noise filtering algorithm) may not yield reasonable results. The moving average is a linear time-invariant low-pass filter, that replaces a data point with the average of neighboring values in a moving window, which smooths out the noise in MCG. Thus, the moving average acts as a low-pass filter, since it removes short-term fluctuations. 
However, the main disadvantage of linear low-pass filters is that $1/f$ noise is in the low-pass band, and it cannot be removed without removing the heart MCG features. 
As a solution to this problem, in this paper, we investigate how deep learning can be used to effectively train on the noisy MCG and corresponding ECG as labels, and then execute the trained model (i.e., at the MTJ-based IoT sensor) to obtain the ECG signal for an unseen MCG input with high accuracy. 

\section{Deep Learning Based Proposed Method}
\label{proposal}

For benchmarking our deep learning model, synthetic MCG data are employed. In the following, the generation of synthetic MCG data and the training procedure for noise processing in MCG signal using a unique deep learning architecture are delineated. We prepared the labeled training MCG data from a public ECG dataset, as outlined in Fig.~\ref{fig:mcggen}. Then, we constructed the deep learning model, consisting of an input layer, a one-dimensional (1-D) convolution layer, a GRU layer, a fully-connected (feed forward) layer, and an output layer as depicted in Fig.~\ref{subfig:cnn_model}. Our deep learning architecture, to identify the underlying ECG signal from the noisy MCG signal, contains 1-D convolution layer to extract the features, and GRU to learn the time-series features.  
After the data collection and pre-processing, the MCG and original ECG signals are used to train the deep learning model depicted in Fig.~\ref{fig:proposed-model}. In the remainder of this section, we describe the details of input labeling, the 1-D convolution, GRU, and fully-connected layers of our proposed model (Fig.~\ref{subfig:cnn_model}), respectively.

\subsection{Labeled Input Preparation}
\label{data-preparation}

\begin{figure}[t!]
	\includegraphics[width=\columnwidth]{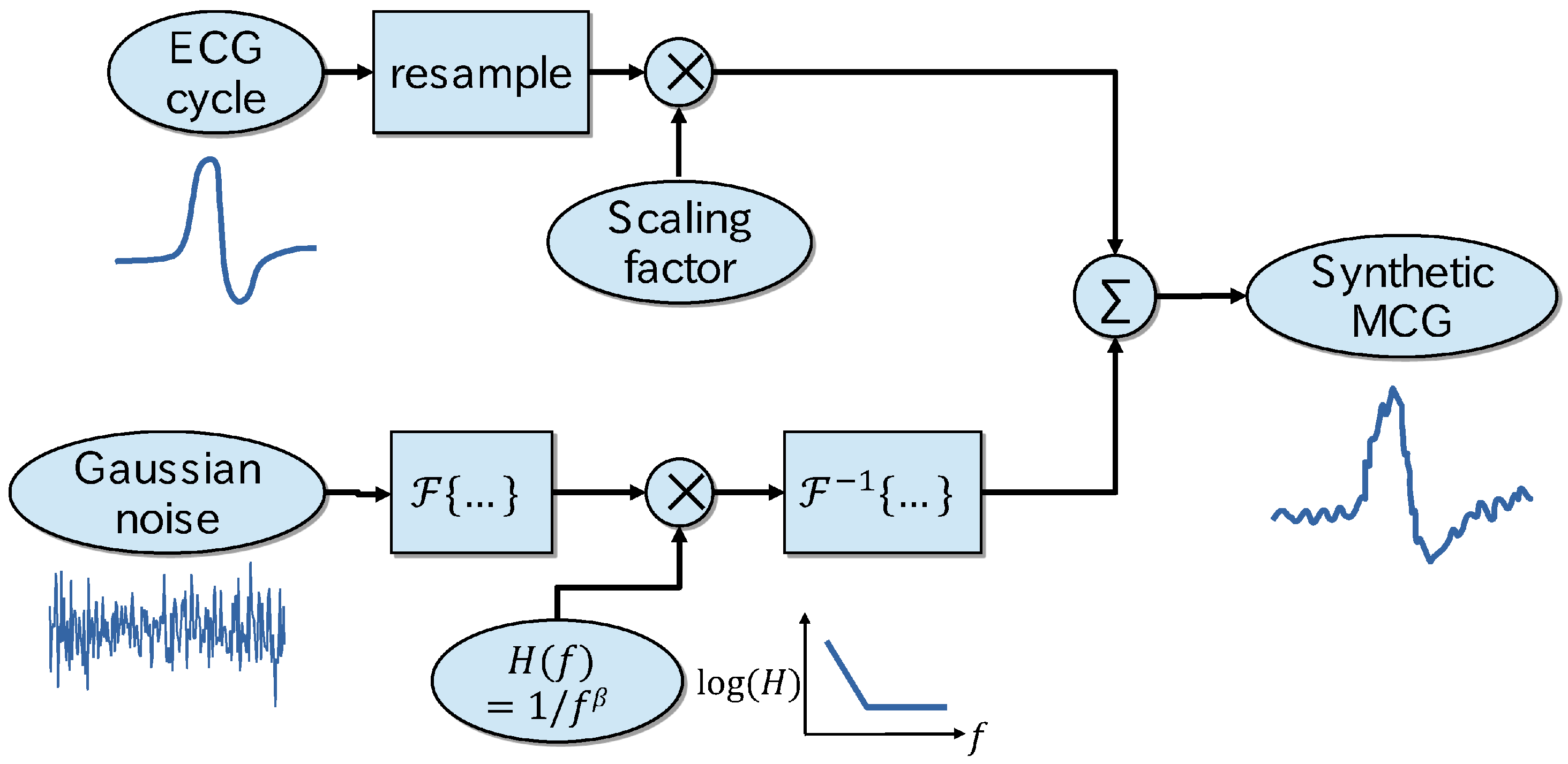}
	\caption{The block diagram of MCG synthesis from ECG cycles.
	}

		\label{fig:mcggen}
\end{figure}

\begin{figure}[t!]
    \centering
          \begin{subfigure}[t]{0.22\columnwidth}
          \centering
        \includegraphics[width=\textwidth]{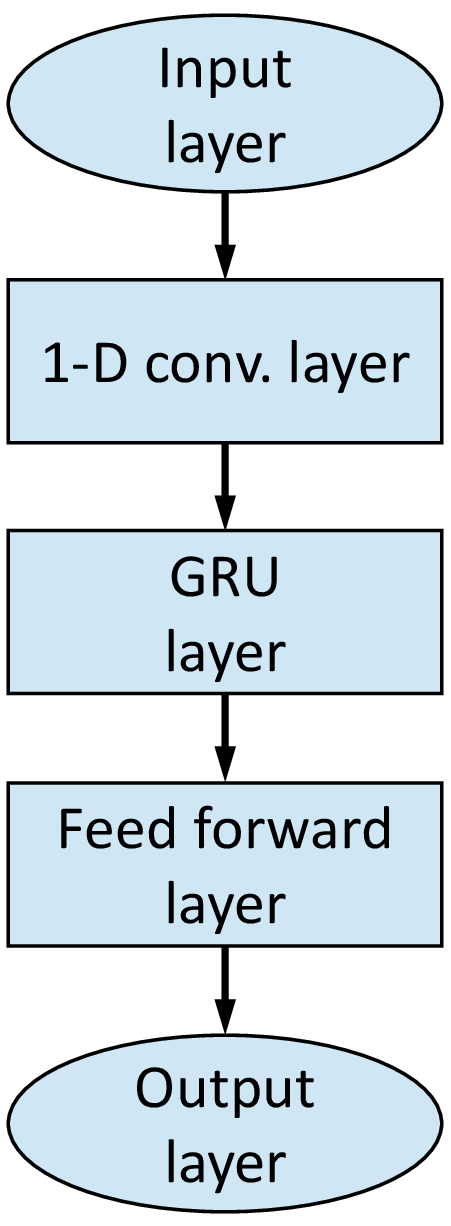}
        \caption{Proposed deep learning model.}
                    \label{subfig:cnn_model}
    \end{subfigure}
        ~\quad
     \begin{subfigure}[t]{0.4\columnwidth}
     \centering
        \includegraphics[width=\textwidth]{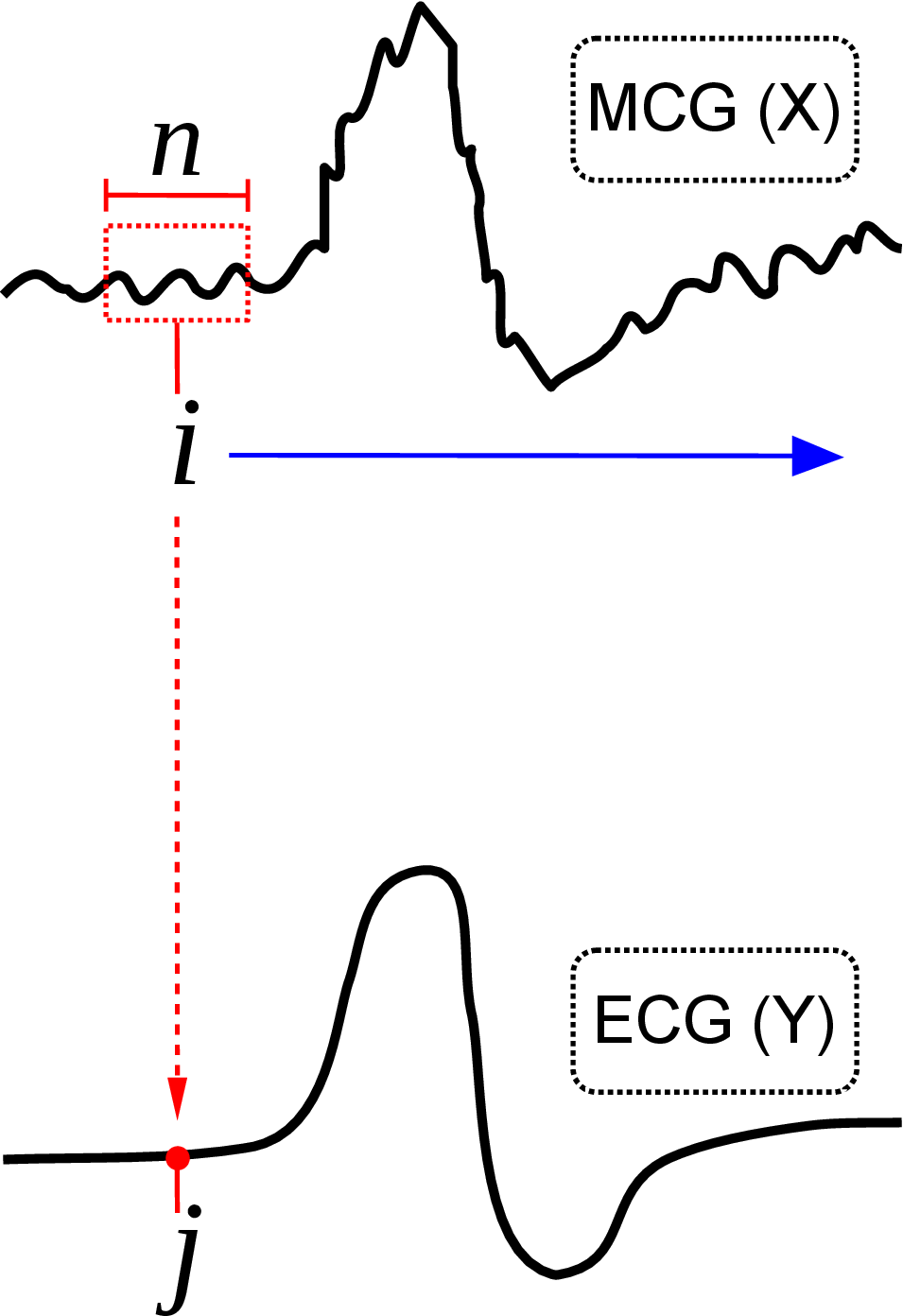}
                \caption{Proposed input/output characterization.}
        \label{subfig:cnn_scheme}
    \end{subfigure}
    \caption{Proposed deep learning model combining the advantages of convolution, GRU, and fully-connected (feed forward) layers for denoising MCG data.}
    \label{fig:proposed-model}
\end{figure}

We synthesized MCG cycles from ECG cycles available in the open PTB Diagnostic Database~\cite{Bousseljot2009, Goldberger_2003}. 
Fig.~\ref{fig:mcggen} demonstrates the synthesis procedure for MCG cycles. We used the cycles from lead II of the healthy individuals provided by Kachuee\textit{et al.}~\cite{Kachuee2018}, which were divided into single heartbeat rhythm cycles starting from $R$ wave to the next $QRS$ complex, with the following sequence ($RSTPQRS$). The cycles were originally sampled at 125 Hz with varying sample lengths and padded zeros. We removed the padded zeros and re-sampled each cycle to a fixed length of 3008, which is approximately 16 times of the original length at a sampling frequency ($f_s$) of 2000 Hz. This resampling was required to accommodate the full spectral features of the MCG noise, without padded zeros to a convolutional layer input. The preconditioned ECG cycle is added to randomly-generated low-frequency noise. The noise in MCG cycles was described based on the expected characters known from real measurements using MTJ sensors \cite{Fujiwara_2018}. We generated the low-frequency noise from a Gaussian white noise with a constant PSD $= 10^{-18} V^2/\text{Hz}$. After transforming the white noise to the frequency domain by Fourier transform $\mathcal{F}$, we applied a transfer function of $1/f^ \beta $ character, then converted the result back into time domain by inverse Fourier transform $\mathcal{F}^{-1}$. The transfer function $H(f)$ was defined as follows:
 \begin{equation}
 H(f) =
  \begin{cases}
        1 & f = 0, \\
        \left(f_k/f\right)^\beta & 0 < f \leq f_\text{k}, \\
        1         & f >    f_\text{k}, 
  \end{cases}    
 \end{equation}
 where the transition between $1/f$ and white noises is set at $f_\text{k} = 250 ~ \text{Hz} = 0.125 f_s$. As a training set, for each ECG cycle, we synthesized 100 MCG cycles with different noise sequences. 
 After the data collection and pre-processing, the MCG and original ECG cycles are used to train the deep learning model depicted in Fig.~\ref{fig:proposed-model}. 
 Instead of using an entire MCG cycle as input to the 1-D convolution layer, it is split into smaller segments, each with sample-size $n$ (as depicted in Fig.~\ref{subfig:cnn_scheme}) with a pre-processing stride length of $\delta$. In the training phase, the MCG segments are not selected in a sequential manner to avoid potential overfitting. On the other hand, in the running/inference phase, the MCG segments are input to the 1-D convolution layer sequentially to obtain real-time mapping of a given MCG segment to its corresponding ECG sample.

\subsection{1-D Convolution Layer}



The input to our considered 1-D convolution layer is the noisy MCG segment.  
The structure extracts noisy MCG information using a number of filters. Each filter is moved across the input using a pre-defined striding length. 
This transforms the input MCG segment into representative features. In other words, the convolution operation preserves the spatial relationship of the MCG segment by learning useful features using stride over the input data. 

Suppose that the input layer consists of $N \times 1$ neurons, followed by our considered 1-D convolution layer with a stride length of one. If we use $m \times 1$ filter $w$, the convolution layer output will be of size $(N-m+1) \times 1$. The $i^{th}$ output of the convolution layer, denoted by $y_i$, can be represented as the activation function of the weighted sum as follows,
\begin{equation}
   y_i = \sigma(  \sum_{a=1}^m w_a x_{(i+a)} ),
\end{equation}
where $x$ denotes the input MCG segment and $\sigma$ refers to the activation function. This convolution operation is carried out using a number of filters, $L$ on the input MCG segment. The output of the 1-D convolution layer, i.e., the generated feature sets, will be of dimension $((N-m+1) \times L)$. 


\subsection{GRU Layer}

The generated feature sets by the 1D-convolution layer are passed as input to a GRU layer of $L$ units~\cite{Le2019}. GRU is an improved variant of the Recurrent Neural Network (RNN). The GRU layer is considered in conjunction with the 1-D convolution layer to deal with the problem of ``vanishing gradient'' and efficiently learn long-term dependencies information of the noise and MCG segment. The GRU layer consists of an update gate and a reset gate. For each element in the input sequence of the GRU layer, the update gate helps the model to estimate how much of the previous information from earlier time-steps needs to be passed along to the future. The update gate $z_t$ for time-step $t$ can be estimated as follows:
\begin{equation}
    z_t = \sigma '(W'_{iz}x'_t + b_{iz} + W'_{hz}h_{(t-1)} + b_{hz}, 
\end{equation}
where $x'_t$ denotes the input at time $t$,  $h_{(t-1)}$ indicate the hidden state of the previous layer at time ($t-1$) or the initial hidden state at time 0, $\sigma '$ is the recurrent activation function, $W'$ refers to the weight, and $b$ is the bias.

On the other hand, the reset gate of the GRU layer decides how much of the past information can be ignored as follows:
\begin{equation}
    r_t = \sigma '(W'_{ir}x_t + b_{ir} + W'_{hr}h_{(t-1)} + b_{hr}).
\end{equation}
The current memory content of the GRU layer utilizes the reset gate to store the relevant information from the past using the following expression:
\begin{equation}
    n_t = \sigma ''(W'_{in}x_t + b_{in} + r_t (W'_{hn}h_{(t-1)} + b_{hn})),
\end{equation}
where $\sigma ''$ is the activation function. 

In the last step, the GRU layer needs to calculate $h_t$ vector, which contains information for the current unit. This is performed using the update gate as follows:
\begin{equation}
    h_t = (1-z_t) n_t + z_t h_{(t-1)}.  
\end{equation}

Thus, the output of the GRU layer is a matrix constructed from $h_t$ vectors of $L$ GRU-units. The dimension of the matrix is $(N-m+1) \times L$.  


\subsection{Fully-connected and output layers}
By flattening the output of the GRU layer, the input to the fully-connected layer (feed forward neural network) is first generated. Backpropagation method~\cite{8450475} is employed in the fully connected layer to adjust the weights of the connections. The fully-connected layer outputs the ECG sample (i.e., a single data point) corresponding to the MCG segment as illustrated in the input/output characterization in Fig.~\ref{subfig:cnn_scheme}.

\begin{figure}[t!]
    \centering
	\includegraphics[width=\columnwidth]{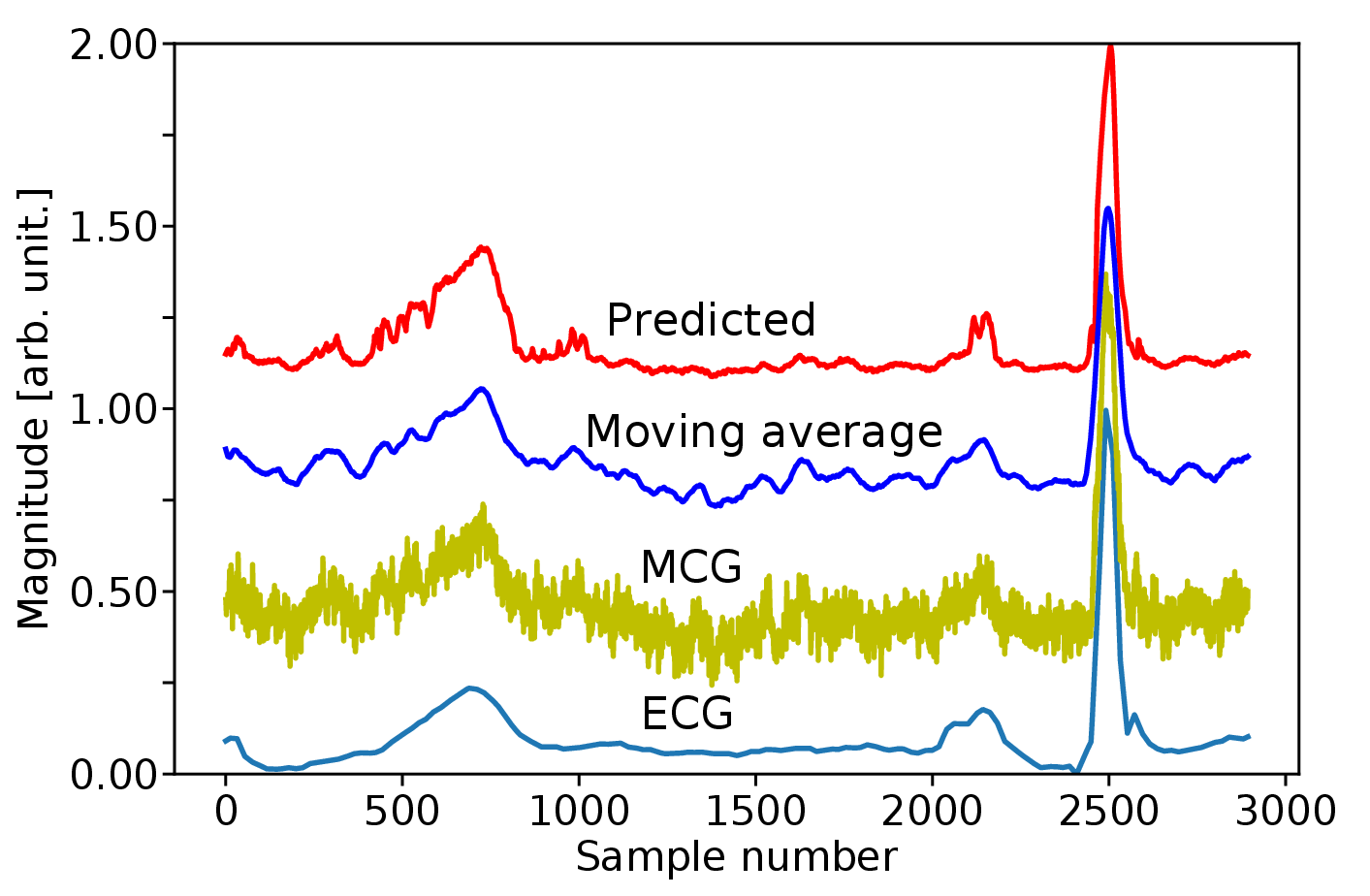}
	\caption{Performance evaluation demonstrating the original ECG cycle, synthetic noisy MCG cycle used as input, and comparison between conventional moving average method and proposed deep learning based method to process and remove the noise from the input signal. The curves are vertically shifted for clarity.}
	\label{fig:cnn_performance}
\end{figure}

\section{Performance Evaluation}
\label{performance}

In order to evaluate the performance of our proposal, we compared our proposed deep learning-based approach against a moving average filter, using data pre-processing and simulations conducted with Python 3 libraries (i.e., NumPy, Matplotlib and Scikit-learn packages) for data processing and visualization. The model is implemented in TensorFlow v1.13 with Keras library for Python.
We conducted training of our proposed deep learning model using 2353 MCG segments per cardiac cycle. 
The pre-processing striding window-size, $\delta$ is set to one while the input MCG segment size is considered to be $N=50$ samples. In the 1-D convolution layer, the number of filters is set to $L=300$, each of which has a size $m=20$. Rectified Linear Unit (ReLU) is used as the activation function ($\sigma$) of the 1-D convolution layer. In the GRU layer, which consists of three hundred units, sigmoid and tanh functions are used as the recurrent activation function ($\sigma '$) and activation function ($\sigma ''$), respectively. The fully-connected layer consists of one hundred neurons with ReLU activation function. The output layer consists of a single neuron with a linear activation function.

Fig.~\ref{fig:cnn_performance} illustrates how our proposal predicts the corresponding ECG cycle from a noisy MCG cycle, and compares it with the conventional moving average method. The original ECG cycle (without noise) is shown at the bottom which is obtained from the original dataset~\cite{Bousseljot2009, Goldberger_2003}. The second curve shows the MCG cycle (with noise), which is synthesized in the data preparation step discussed in Sec.~\ref{data-preparation}. The third curve demonstrates the MCG cycle obtained after applying the moving average filter with sliding window-size of 50. The fourth curve shows the outcome of our proposed deep learning model, i.e., the predicted ECG cycle (without noise) based on the noisy MCG input cycle. Comparing both the predicted and average-filtered cycles with the original ECG segment in Fig.~\ref{fig:cnn_performance}, we notice that both moving average and deep learning techniques are able to construct the underlying ECG cycle from the noisy input MCG cycle. But at times, the deep learning model distinguishes the low-frequency noise features from intrinsic ECG features, \emph{e.g.}~in the samples interval 1000--2000, without degrading the $QRS$ features that have higher frequency components.

\begin{figure}[t!]
        \centering
        \begin{subfigure}[t]{\columnwidth}
                \centering
                \includegraphics[width=\columnwidth]{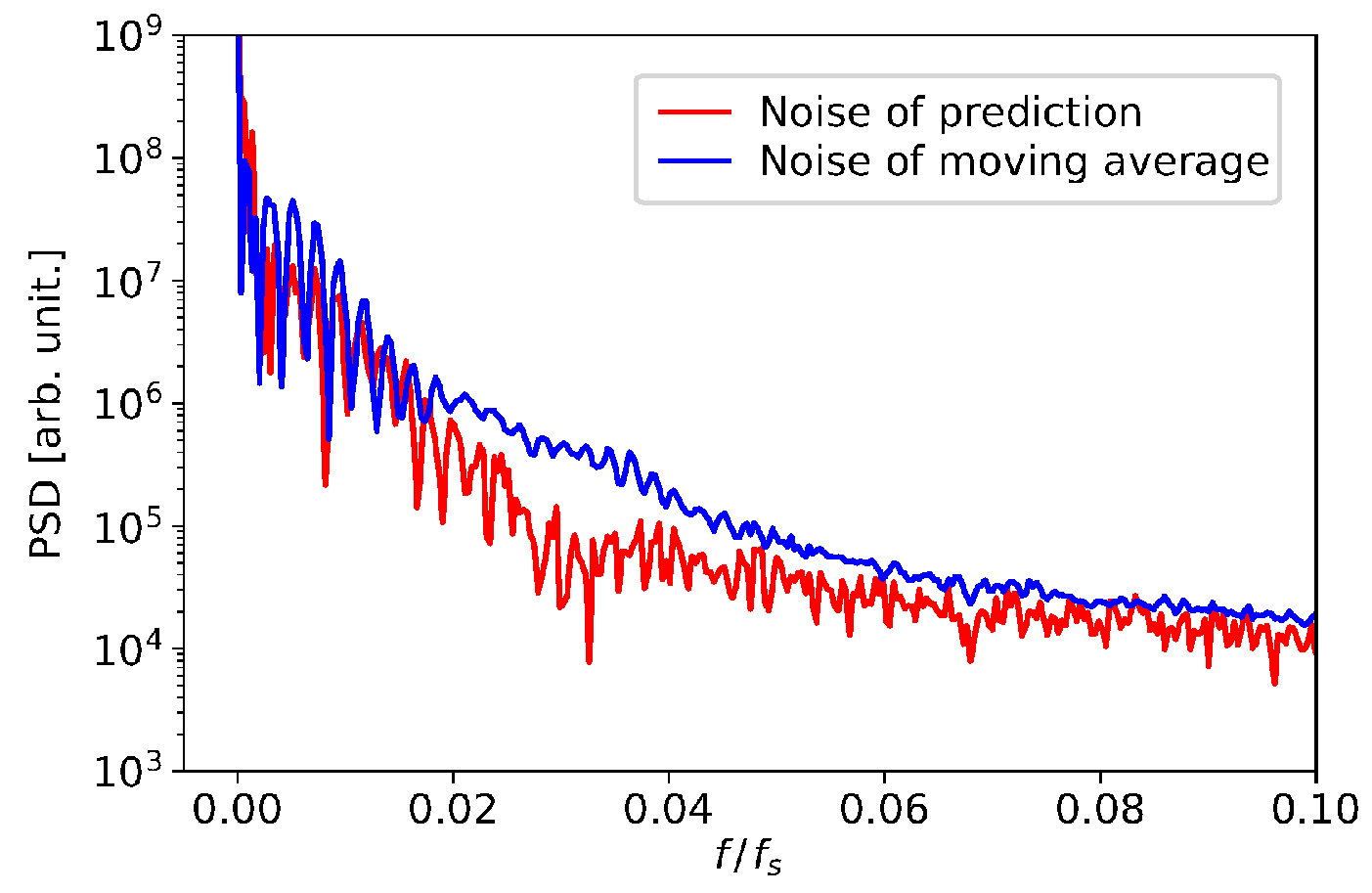}
        \vspace{-0.4cm}
	            \caption{Dependence of noise power on spectral frequency for the deep learning prediction and the moving average filtering. The spectral frequency is normalized by the sampling frequency $f_s$.
	            }
	            \label{fig:PSD2}
        \end{subfigure}%
        
        \begin{subfigure}[t]{\columnwidth}
               \centering \includegraphics[width=\columnwidth]{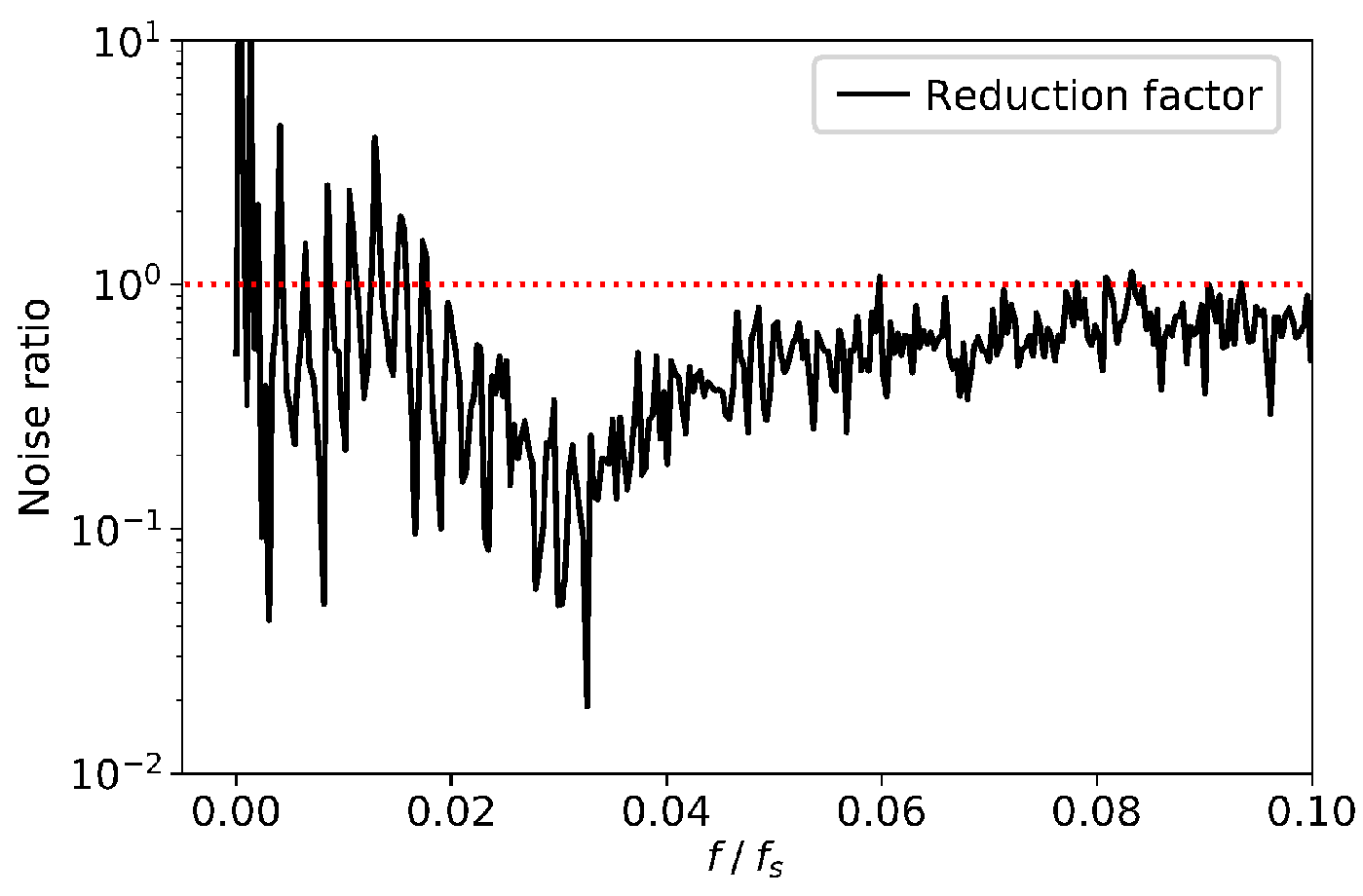}
	            \caption{Noise PSD ratio of deep learning prediction noise relative to the noise of the moving average filter. The dashed line is the unity ratio.
	            }
	            \label{fig:noise-ratio}
        \end{subfigure}%
        ~
        
        \caption{The noise power spectral density of the deep learning prediction is lower than that of moving average filtering, especially at low frequencies.
        }\label{fig:spectral}
        \vspace{-0.1cm}
\end{figure}
In order to investigate the performance of the deep learning model, we compare the spectral features of the remaining noise after applying each method of moving average and our deep learning-based prediction. We plot the Power Spectral Density (PSD) dependence on spectral frequency normalized by the sampling frequency ($f/f_s$) in Fig.~\ref{fig:PSD2}. We averaged the remaining noise PSD in the output of prediction and moving average filters of 
unseen 2,000 noisy MCG cycles. Both the filtering methods decreased noise and lowered the $1/f$ knee frequency from its original value at $f_k/f_s = 0.125$. However, at lower frequencies, the noise power of our proposed deep learning-based prediction model remains lower than that of the moving average filter. The effect of this is highlighted further in Fig.~\ref{fig:noise-ratio}, which plots the noise reduction factor for the prediction noise and moving average noise. Our deep learning-based model outperforms the moving average at frequencies ranging $f/f_s \in \left[ 0.02, 0.05\right]$. This frequency band has major MCG spectral components, resulting from $QRS$ complex. As expected, the spectral resolution of the pre-processing striding window is the inverse of the window-size, which is $1/50 = 0.02$. This means that the deep learning-based model performs the best at discerning noise that lasts between half and the full window size. Further adjustments to the pre-processing striding window-size can be used to tune out the low-frequency noise without affecting the MCG features.

\section{Conclusion}
\label{conclusion}

Since ECG cannot be applied effectively for remote/home monitoring of cardiac condition of patients, new applications with new values in the IoT industry for cardiovascular monitoring need to be considered. In this regard, spintronic sensors using Magentic Tunnel Junction (MTJ) devices offer a strong advantage in terms of high sensitivity and portability as well as its ability to facilitate logic-in-sensor architecture. 
The paper also addressed the key challenge of dealing with the intrinsic $1/f$ noise at the sensor, where the linear filtering methods (such as moving average) cannot effectively remove it due to similar low-frequency characteristics with the heart signal. A deep learning-based model, combining the advantages of 1-D convolution, GRU, and fully-connected (feed forward neural network) layers, is used to process this noise. The model was trained with MCG data synthesized from a public ECG dataset. The results indicated encouraging performance of our proposed method compared with the conventional moving average technique. The noise power in the model predictions showed a large reduction at low frequencies compared to the moving average technique. Therefore, by more optimization of the deep learning model, the smart removal of low-frequency noise can unlock the potential for MCG monitoring applications using the MTJ IoT sensors. Furthermore, in the future, this noise processing can be performed on-node within the logic-in-sensor architecture.

\balance



\balance

\section*{Acknowledgments}  
This work was partially supported by the Center for Science and Innovation in Spintronics (Core Research Cluster), Center for Spintronics Research Network, Tohoku University, the S-Innovation program, Japan Science and Technology Agency (JST), and by JSPS KAKENHI Grant Number JP19K15429.
The authors would like to thank Dr. Kenji Mizuguchi for his support.
The statements made herein are solely the responsibility of the authors.

\bibliographystyle{IEEEtran}

\bibliography{refs}

\end{document}